\documentclass[10pt,twocolumn,letterpaper]{article}

\usepackage[pagenumbers]{cvpr} 
\usepackage[utf8]{inputenc}

\usepackage[dvipsnames]{xcolor}
\usepackage{booktabs}
\usepackage[
  separate-uncertainty=true,
  multi-part-units=single,
  list-units=repeat,
  detect-weight,
  mode=text
]{siunitx}

\usepackage{lipsum}

\usepackage{tikz}
\newcommand{\tikzcircle}[1][fill=red]{\tikz[baseline=-0.625ex]\draw[white,#1,radius=3pt] (0,0) circle ;}

\newcommand{\legend}{\protect{(\tikzcircle[fill=Gray]) Control, (\tikzcircle[fill=Cerulean]) $0.06\times$, (\tikzcircle[fill=BurntOrange]) $0.32\times$, (\tikzcircle[fill=LimeGreen]) $1.60\times$, (\tikzcircle[fill=RedOrange]) $8.00\times$, and (\tikzcircle[fill=Orchid]) $40.0\times$}}

\definecolor{mygreen}{HTML}{66c2a5}
\definecolor{myorange}{HTML}{fc8d62}
\definecolor{mypurple}{HTML}{8da0cb}
\definecolor{mypink}{HTML}{e78ac3}
\newcommand{\legendary}{\protect{(\tikzcircle[fill=mygreen]) Site 1, (\tikzcircle[fill=myorange]) Site 2, (\tikzcircle[fill=mypurple]) Site 3, (\tikzcircle[fill=mypink]) Site 4}}

\newcommand\sbullet[1][.5]{\mathbin{\vcenter{\hbox{\scalebox{#1}{$\bullet$}}}}}
\newcommand\ssquare[1][.5]{\mathbin{\vcenter{\hbox{\scalebox{#1}{$\blacksquare$}}}}}

\definecolor{cvprblue}{rgb}{0.21,0.49,0.74}
\usepackage[pagebackref,breaklinks,colorlinks,citecolor=cvprblue]{hyperref}

\def\paperID{*****} 
\def\confName{CVPR}
\def\confYear{2024}

\title{Grad-CAMO: Learning Interpretable Single-Cell Morphological Profiles from 3D Cell Painting Images}
\author{
    Vivek Gopalakrishnan \\
    Xellar Biosystems and MIT \\
    {\tt\small vivekg@mit.edu}
\and
    Jingzhe Ma \\
    Xellar Biosystems \\
    {\tt\small jma@xellarbio.com}
\and
    Zhiyong Xie \\
    Xellar Biosystems \\
    {\tt\small zxie@xellarbio.com}
}

\begin{document}
\maketitle

\begin{abstract}
Despite their black-box nature, deep learning models are extensively used in image-based drug discovery to extract feature vectors from single cells in microscopy images. To better understand how these networks perform representation learning, we employ visual explainability techniques (\eg Grad-CAM). Our analyses reveal several mechanisms by which supervised models cheat, exploiting biologically irrelevant pixels when extracting morphological features from images, such as noise in the background. This raises doubts regarding the fidelity of learned single-cell representations and their relevance when investigating downstream biological questions. To address this misalignment between researcher expectations and machine behavior, we introduce Grad-CAMO, a novel single-cell interpretability score for supervised feature extractors. Grad-CAMO measures the proportion of a model's attention that is concentrated on the cell of interest versus the background. This metric can be assessed per-cell or averaged across a validation set, offering a tool to audit individual features vectors or guide the improved design of deep learning architectures. Importantly, Grad-CAMO seamlessly integrates into existing workflows, requiring no dataset or model modifications, and is compatible with both 2D and 3D Cell Painting data. 
Additional results are available at \url{https://github.com/eigenvivek/Grad-CAMO}.
\end{abstract}

\section{Introduction}
Drug discovery, screening, and development is a lengthy and costly process with a high failure rate, due in large part to the numerous discrepancies between cells cultured on 2D surfaces and those inhabiting \textit{in vivo} cellular environments~\cite{langhans2018three}. 3D tissue culture techniques, particularly microfluidic-based organ-on-a-chip platforms~\cite{bai2024rapid}, enable the development of cellular disease models that more closely mimic \textit{in vivo} biology. When combined with high-content, high-throughput imaging techniques, such as the Cell Painting assay~\cite{bray2016cell}, these advanced tissue culture approaches can be used to measure the effect of diverse treatments on 3D cellular morphology. However, for cellular analyses to scale with the immense rate with which modern imaging platforms acquire data, we also require computational methods that automatically transform single cells in 3D Cell Painting images into quantitative representations.
\begin{figure*}[t]
    \centering
    \includegraphics[width=\linewidth]{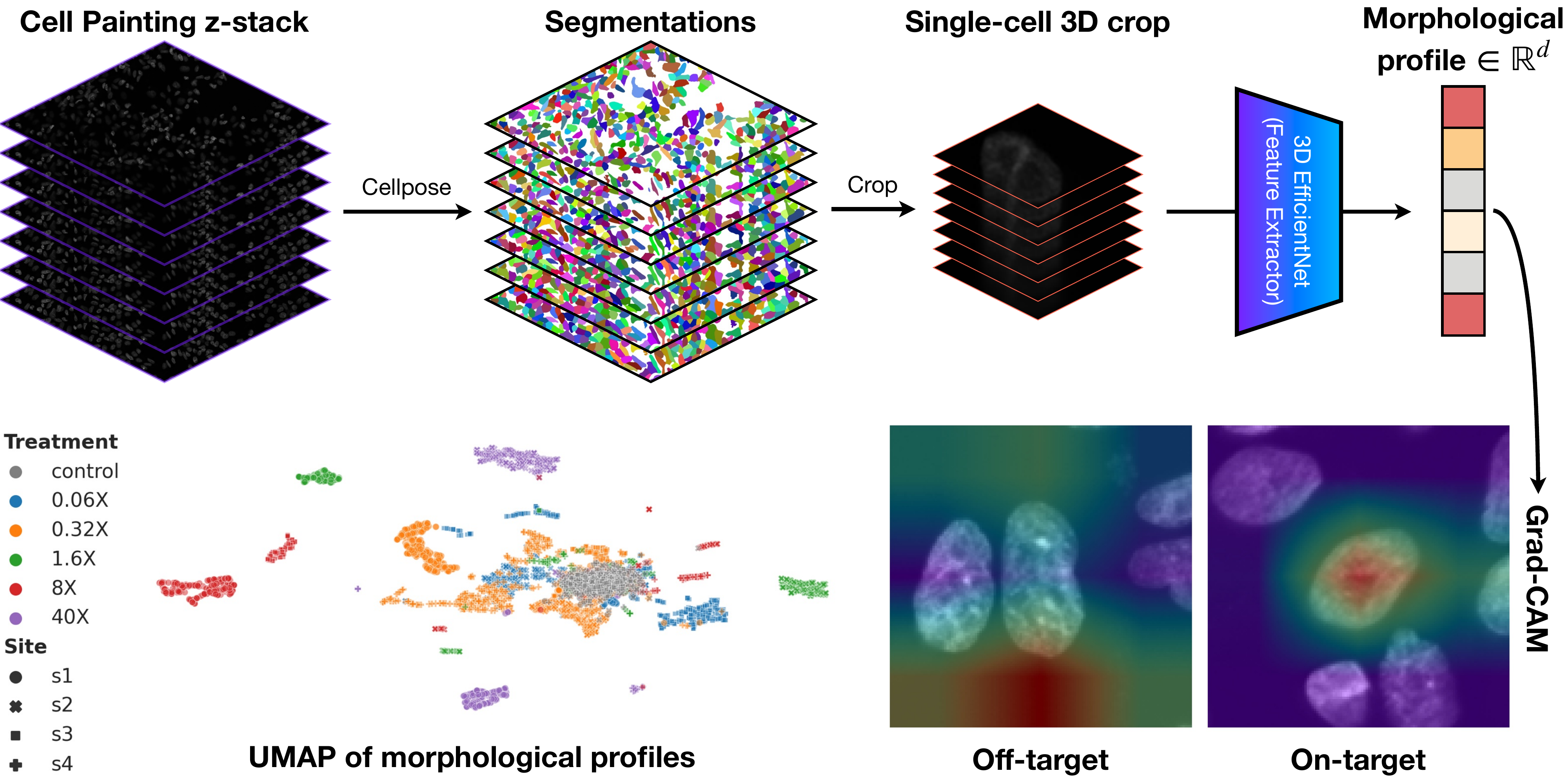}
    \caption{\textbf{Overview of Grad-CAMO.} Given a 3D Cell Painting Z-stack, we first use Cellpose~\cite{stringer2021cellpose} to segment individual cells. Using segmentation masks, we create cuboid 3D crops centered on single cells. Extending approaches commonly used for 2D Cell Painting images (\eg DeepProfiler~\cite{moshkov2024learning}), we train a 3D EfficientNet~\cite{tan2019efficientnet} to predict the treatment label of an individual cell from a crop. During inference, held out cells are passed to the trained network and activations at intermediate layers are used as single-cell feature vectors in $\mathbb R^d$. Visualization of feature vectors using UMAP~\cite{mcinnes2018umap} shows single cells are highly separable based on learned morphological profiles. However, interpretability analysis using Grad-CAM demonstrates that deep learning-based feature extractors do not always pay attention to the cell-of-interest when forming single-cell morphological profiles (on-target \vs off-target). To quantify the fidelity of learned morphological profiles, we introduce \textbf{Grad-CAMO}, a single-cell interpretability metric to quantify the level of confounding in a model's predictions.}
    \label{fig:abstract}
\end{figure*}

Quantifying the phenotypic effects of experimental perturbations from high-throughput imaging assays, a feature extraction process known as morphological profiling, is a necessary and challenging step in the analysis of Cell Painting data. Whereas traditional methods use handcrafted image processing algorithms to extract human-designed descriptors of cellular morphology (\eg CellProfiler~\cite{carpenter2006cellprofiler}), recent methods have taken a deep learning approach, training a neural network to extract features learned to be relevant from raw imaging data~\cite{moshkov2024learning, moen2019deep}. While learned morphological profiles have been shown to enable better performance in downstream analysis tasks, the question of interpretability limits deep learning-based approaches: how can we ensure that the morphological profiles extracted by black-box deep learning models capture biologically relevant information about single cells, rather than simply exploiting confounding factors present in image data (\eg batch effects) to minimize training loss? 

Previous work has suggested that pretraining supervised feature extractors on large, diverse sets of 2D Cell Painting images can mitigate the impact of confounders on learned morphological profiles~\cite{moshkov2024learning}. Unfortunately, for 3D Cell Painting, there are no open-source imaging databases as extensive as the JUMP-Cell Painting or Recursion datasets~\cite{chandrasekaran2023jump, sypetkowski2023rxrx1}, which contain millions of 2D microscopy images for pretraining deep learning models. This data limitation also precludes the development of self-supervised feature extractors (\eg attention-based mechanisms such as DINO~\cite{caron2021emerging}), which have shown promise in 2D Cell Painting images~\cite{pfaendler2023self}. As transformer-based models are currently infeasible in for 3D microscopy data, there remains a need for quantitative methods to audit the morphological profiles produced by supervised deep learning models.

To address this need, we propose evaluating learned morphological profiles with Gradient-weighted Class Activation Mapping (Grad-CAM)~\cite{selvaraju2017grad}, a technique that uses the gradients of a convolutional neural network to localize which regions of an input image the model paid the most attention to when making its prediction. 
When combined with single-cell segmentations masks, we can measure the proportion of a model's attention that overlaps with the cell of interest compared to the background. 
This interpretability metric, which we term Grad-CAMO (\ie Grad-CAM Overlap), enables identification of morphological profiles that faithfully describe biologically relevant components of the input image, helping to quantify the influence of confounders on the extracted features. 

In our experiments, we demonstrate the utility of Grad-CAMO for 3D Cell Painting images: using a dataset of single-cell 3D crops extracted from fluorescence microscopy Z-stacks, we train a 3D convolutional neural network to predict the treatment label of each cell. Intermediate activations are extracted and used as single-cell features. Using Grad-CAMO, we find that only 30\% of learned morphological profiles have Grad-CAM localization maps that meaningfully overlap with the cell's segmentation map. Visualization of the localization maps demonstrates that supervised feature extractors can cheat by exploiting irrelevant, non-biological information in microscopy data. 

\paragraph{Contributions.}
We introduce Grad-CAMO, an easily computed interpretability metric that quantifies to what extent a single-cell morphological profile represents the cell of interest. Like the Grad-CAM localization map, Grad-CAMO is data- and model-agnostic: it can be computed for any convolutional feature extractor without modification to the imaging workflow or deep learning architecture. However, unlike previous interpretability techniques, which only provide a qualitative assessment of fidelity, Grad-CAMO provides a quantitative measure of the biological relevance of the morphological profile for each single cell. This highlights the utility of our metric for screening computational pipelines that process millions of cells from high-throughput imaging data. Although this work primarily focused on 3D images, Grad-CAMO is also extensible to 2D Cell Painting images and can be used to explain descriptions made by existing models employed in academia and industry. The integration of Grad-CAMO into a traditional single-cell feature extraction pipeline is shown in \Cref{fig:abstract}.

\section{Preliminaries}
Let $\mathbf I : \mathbb R^3 \to \mathbb R^C$ represent a Z-stack comprising thousands of cells, where $\mathbf I$ is a function mapping 3D coordinates to pixel intensities in a 3D Cell Painting image with $C$ fluorescent channels. Let $\mathbf V_i \subseteq \mathbf I$ be a single-cell 3D crop of the Z-stack centered on the $i$-th cell. Finally, let $Y_i \in \mathcal Y = \{1, \dots, K\}$ be a label for the $i$-th cell, corresponding to a categorical dimension (\eg the treatment dosage) along which we wish to extract morphological profiles for some downstream task. 
We assume there is at least one Z-stack for each categorical label and that images come from wells wherein each cell received the same treatment.

\subsection{Supervised Feature Extraction}
To extract morphological profiles using supervised learning, we first train a convolutional neural network $f_\theta$ to predict $Y_i$ from $\mathbf V_i$ by minimizing a classification loss with respect to weights $\theta$ over some subset of cells in the Z-stacks. 
Then, at inference time, cells not seen by the network during training are processed by the network and activations at an intermediate layer are extracted as learned morphological profiles.
To use explicit notation, let us decouple $f_\theta$ into a convolutional backbone $g_\varphi$ and a classification head $h_w$ such that
\begin{equation}
    \hat Y_i = f_\theta(\mathbf V_i) = (h_w \circ g_\varphi)(\mathbf V_i) \,,
\end{equation}
where $\hat Y_i$ is the predicted treatment label for cell $\mathbf V_i$.
Then, for a previously unseen cell $\mathbf V$, the intermediate activation
\begin{equation}
    \mathbf A = g_\varphi(\mathbf V) \in \mathbb R^{C' \times H \times W \times D}
\end{equation}
is a representation of the single-cell crop that the network has learned is useful for predicting the treatment label. This representation can then be pooled and reshaped into a feature vector in $\mathbb R^d$. 
Representing 3D Cell Painting images as a set of single-cell morphological profiles enables numerous downstream machine learning tasks such as clustering and classification.

\subsection{Grad-CAM}
While a neural network given sufficient data will eventually learn to accurately estimate the treatment label of a single-cell crop, it is not obvious \textit{how} the network arrived at its prediction. This calls into the question the fidelity of the intermediate activation $\mathbf A$, as it could ostensibly arise from a spurious correlation or confounding variable latent in the input data. To visually explain the decisions made by a convolutional neural network, we can use Gradient-weighted Class Activation Mapping (Grad-CAM)~\cite{selvaraju2017grad}, an interpretability technique that exploits the model's gradients to produce a localization map highlighting which regions in the input image led the model to predict a particular label. Specifically, for a cell that the model predicts received treatment $k \in \mathcal Y$, the Grad-CAM localization map is a linear combination of the channels in $\mathbf A$, calculated as 
\begin{equation}
    \label{eq:grad-cam}
    \Tilde{\mathbf G} = \mathrm{ReLU} \left( \frac{1}{C'} \sum_{c=1}^{C'} w_c^{(k)} \mathbf A_c \right) \in \mathbb R^{H \times W \times D} \,,
\end{equation}
where the weight $w_c^{(k)}$ is computed with  per-channel average pooling:
\begin{equation}
    w_c^{(k)} = \frac{1}{HWD} \sum_{x=1}^H \sum_{y=1}^W \sum_{z=1}^D \frac{\partial Y^{(k)}}{\partial \mathbf A_c(x, y, z)} \,,
\end{equation}
where $Y^{(k)}$ is the model's score for treatment label $k$ (\ie the output prior to applying the softmax). This weight captures the importance of the $c$-th channel to the network's prediction, and the ReLU in \cref{eq:grad-cam} serves to preserve only those regions that positively influence this prediction.

Note that $\Tilde{\mathbf G}$ only provides a coarse localization since it is the same shape as $\mathbf A$, and intermediate activations at bottleneck layers in convolutional neural networks typically have a much smaller spatial dimensions than the input. Therefore, we upsample $\Tilde{\mathbf G}$ to match the original dimension of $\mathbf V$ using trilinear interpolation and refer to the rescaled Grad-CAM localization map as $\mathbf G = \mathrm{Trilinear}(\Tilde{\mathbf G})$.

\section{Methods}

\subsection{Grad-CAMO}
While Grad-CAM can be used to visually inspect the outputs of deep networks, it is infeasible to manually inspect every localization map for the thousands of cells in a Z-stack. Therefore, we introduce Grad-CAMO (short for Grad-CAM Overlap), a single number score to quantify the concentration of the localization map within the cell of interest.
Let $\mathbf M$ be a binary segmentation mask with the same shape as $\mathbf V$ that denotes the occupancy of the central cell within the volume. Let $\mathbf G$ be the localization map produced upsampled to match the dimensions of $\mathbf V$. Then, we can compute the Grad-CAMO score as
\begin{equation}
    \mathrm{Grad\text{-}CAMO}(\mathbf G, \mathbf M) = \frac{\| \mathrm{vec}(\mathbf G \odot \mathbf M) \|_1}{\ |\mathrm{vec}(\mathbf G) \|_1} \,,
\end{equation}
where $\odot$ represents element-wise matrix multiplication. 

Grad-CAMO is a model- and data-agnostic score, able to be automatically computed for any convolutional neural network.
Additionally, although not explored in this work, Grad-CAMO is extensible to vision transformer-based feature extractors. This can be accomplished simply by replacing the Grad-CAM localization map $\mathbf G$ with the self-attention map produced by vision transformer models.

\begin{figure*}[t]
    \centering
    \includegraphics[width=\linewidth]{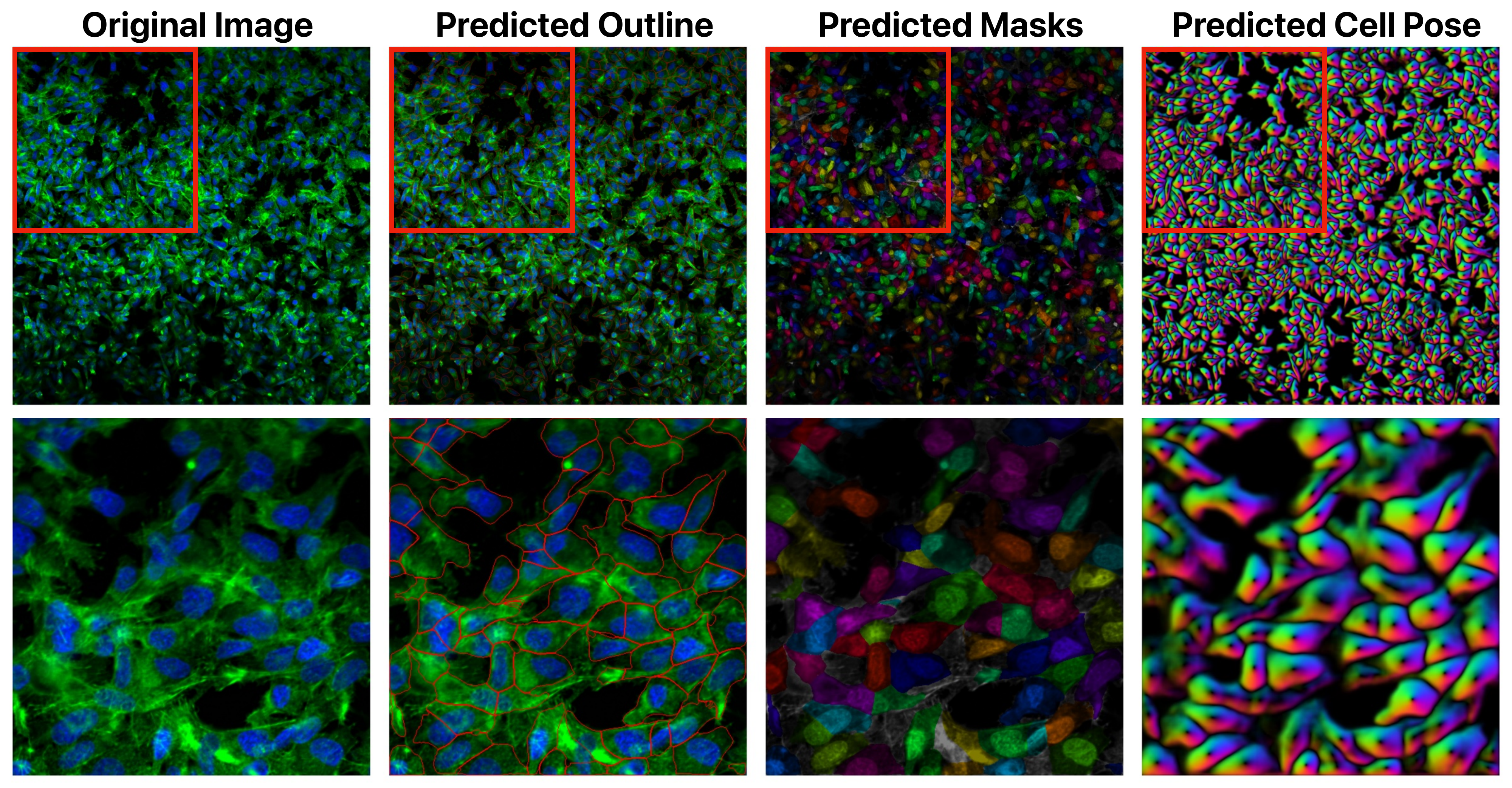}
    \caption{\textbf{Automatic segmentations produced by Cellpose.} Zoomed-in image patches are shown to demonstrate the accuracy of segmentations produced by Cellpose with minimal hyperparameter tuning. 2D segmentation masks were stitched across the $z$-axis of the 3D Cell Painting images to form 3D segmentation masks.}
    \label{fig:cellpose}
\end{figure*}

\subsection{Model Auditing}
The Grad-CAMO score ranges from $[0, 1]$, with higher scores denoting morphological profiles that more closely relate to the cell of interest. Because this score can be automatically evaluated for every cell independent of model architecture, Grad-CAMO can be used to audit supervised feature extractors, either by filtering cells with biologically irrelevant morphological profiles or by providing a heuristic to guide improvements to the design of the feature extractor. For example, let $\mathcal D_{\mathrm{val}} = \{\mathbf V_1, \dots, \mathbf V_N\}$ be a validation set of $N$ held-out single-cell crops. For each cell, we can compute a Grad-CAMO score $\{s_1, \dots, s_N\}$. Optimal design choices (\eg what layer to extract features from, what pooling function to use, \etc) can be empirically determined by maximizing a central measure of Grad-CAMO over the validation dataset, \eg $\hat s = \frac{1}{N} \sum_{i=1}^N s_i$. Thus, Grad-CAMO a useful metric for hyperparameter tuning.

\subsection{Implementation Details}

\paragraph{Single-cell segmentation.}
\label{sec:implementation}
We segment individual cells in 3D Cell Painting images using the \texttt{cyto} model in Cellpose~\cite{stringer2021cellpose}, which only uses the nucleus and cytoplasm channels in the Z-stack. We do not use the built-in 3D segmentation model, but rather achieve higher performance by segmenting every slice in the Z-stack with the 2D model and combining all slices into a 3D segmentation mask using a stitch threshold of 0.05. Additional non-default hyperparameters used include a flow threshold of 0.75, an anisotropy factor of $1.71$ (\ie the ratio of between-plane pixel spacing to in-plane pixel spacing), and a minimum pixel size of 30 pixels per 2D segmentation mask. A 2D slice of the 3D segmentation masks produced by Cellpose is shown in \Cref{fig:cellpose}. 

\begin{figure*}[t]
    \centering
    \includegraphics[width=\linewidth]{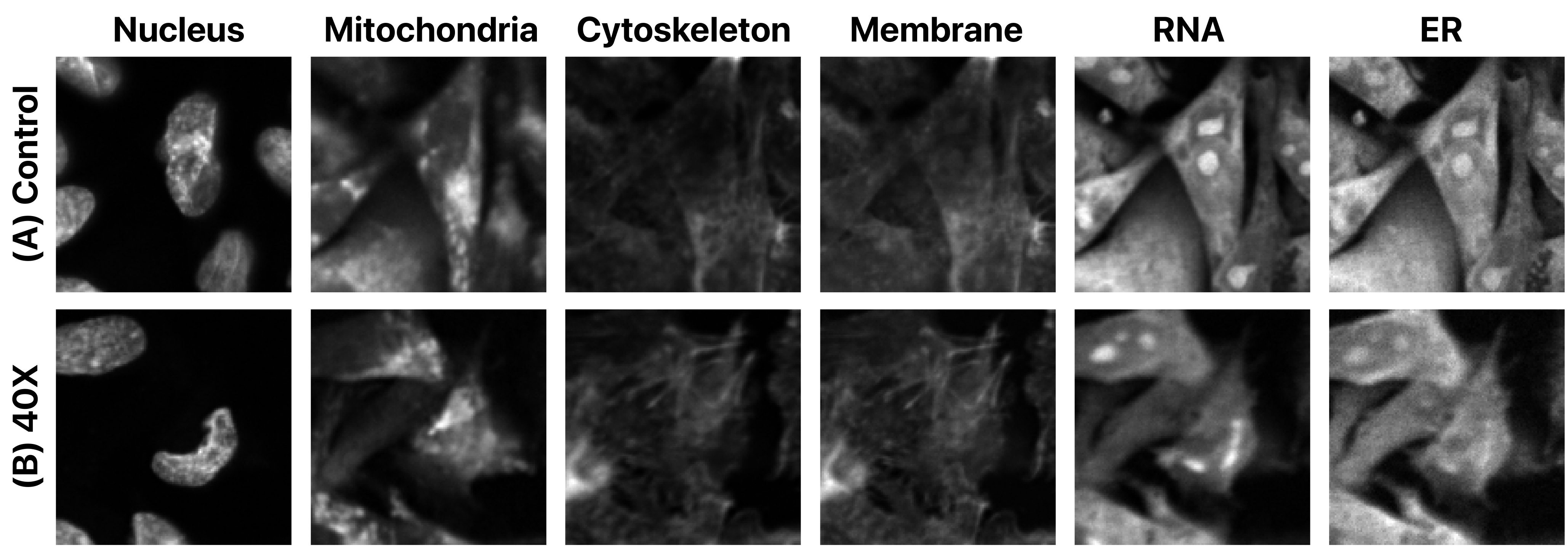}
    \caption{\textbf{Examples of multi-channel single-cell crops at different dosages.} For every cell segmented by Cellpose, a $128 \times 128 \times 21$ crop was extracted from the Z-stack and preprocessed to standardize pixel intensity ranges across samples.}
    \label{fig:crops}
\end{figure*}

Using 3D segmentation masks, $128 \times 128 \times Z$ single-cell crops were extracted from the Z-stack images, where $Z=21$ is the total number of slices in the Z-stack. This was done as nearly all cells spanned the entire $z$-axis. For each crop, the cell center was determined as the average of the minimum and maximum pixel index in the $x$- and $y$-axes of the cell's 3D segmentation mask. 
For preprocessing, channels in each single-cell crop were first rescaled within $[0, 1]$, then normalized by the mean and variance of rescaled pixel intensities in the training set.
A 2D slice of a single-cell crop produced by our segmentation and preprocessing pipeline is shown in \Cref{fig:crops}.

\paragraph{Supervised learning.}
We trained a 3D EfficientNet-B0 to predict the treatment label of single-cell crops.\footnote{We used an open-source implementation available at \url{https://github.com/shijianjian/EfficientNet-PyTorch-3D}.}
To ensure that single-cell crops were divisible by the number of pooling layers in the model, we upsampled all 3D volumes from their original size of $128 \times 128 \times 21$ voxels to $224 \times 224 \times 64$ voxels using trilinear interpolation.
After normalization, all single-cell crops were augmented as follows:
\begin{enumerate}
    \item \textbf{Random $x$- and $y$-axis flips:} in our implementation, the $z$-axis is never reversed.
    \item \textbf{Random brightness adjustment:} add a random constant intensity from $\mathcal U[0.5, 1.25]$ to all pixel intensities.
    \item \textbf{Random gamma adjustment:} raise all pixel intensities to a random power sampled from $\mathcal U[0.5, 1.5]$.
\end{enumerate}
We do not include random crop and resize augmentations as they alter cell size, a biologically relevant feature that can provide information about the health of a cell under different treatments.
All augmentations were applied randomly with probability $p=0.5$. During training, we minimized cross entropy loss using the Adam optimizer with a learning rate of $3 \times 10^{-4}$. All models were trained on a single NVIDIA A6000 GPU with a batch size of 8 single cells.

\paragraph{Correcting for batch effects.}
We employ the whitening transform fit to the control cells as our batch effect correction strategy.
In statistical terms, if $\mathbf x \in \mathbb R^{d}$ is a random feature vector with mean $\mathbf 0$ and covariance $\mathbf \Sigma$, the matrix $\mathbf W \in \mathbb R^{d \times d}$ performs the whitening transform if $\mathbf W \mathbf x$ has a covariance matrix equal to identity. That is, the whitening transform maps samples from a particular distribution to isotropic Gaussian noise. 
Given a (centered) data matrix $\mathbf X \in \mathbb R^{N \times d}$ whose rows are feature vectors from $N$ control cells, we estimate $\mathbf W = \hat{\mathbf \Sigma}^{-\frac{1}{2}}$ via an eigendecomposition of the sample covariance matrix $\hat{\mathbf \Sigma} = \frac{1}{N} \mathbf X^T \mathbf X$~\cite{moshkov2024learning}.
Cells in the control group are used to estimate the whitening transform as they should have identical phenotypic profiles, and thus, any differences in their features are sampled from the underlying noise distribution. Given feature vectors $\mathbf Y \in \mathbb R^{M \times d}$ from other cells, we use the batch-corrected features $\mathbf W \mathbf Y^T$.

\section{Results}

\subsection{Dataset}
Our 3D Cell Painting dataset was derived from a drug-induced liver injury study of Hep G2 cells treated with Stavudine at six different concentrations. Cells were seeded in a 3D growth medium and cultured for 14 days. Two wells were cultured at each concentration, and four distinct sites were imaged per well. Cells were stained with six fluorescent dyes including Hoescht (nucleus), Concanavalin (endoplasmic reticulum), SYTO14 (nucleoli and cytoplasmic RNA), Wheat-Germ Agglutinin (cell membrane and Golgi apparatus), Phalloidin (F-actin cytoskeleton), and MitoTracker (mitochondria)~\cite{bray2016cell}. 3D Cell Painting data were acquired using six channels at 20$\times$ magnification and an anisotropic voxel spacing of $\SI{0.3505}{\micro\m} \times \SI{0.3505}{\micro\m} \times \SI{0.6}{\micro\m}$. For each volume, 21 slices were acquired along the $z$-axis. 

\begin{table}[b]
\centering
\caption{\textbf{3D Cell Painting dataset.} Six treatment groups were cultured with two biological replicates and four technical replicates. Cell counts were estimated using segmentations from Cellpose.}
\label{tab:dataset}
\begin{tabular}{@{}ccc@{}}
\toprule
\textbf{Treatment Dose} & \textbf{Well} & \textbf{Number of Cells} \\ \midrule
Control & D02, D03 & 9,313 \\
0.06$\times$ & B02, B03 & 4,119 \\
0.32$\times$ & C02, C03 & 10,182 \\
1.60$\times$ & G02, G03 & 9,468 \\
8.00$\times$ & F02, F03 & 9,813 \\
40.0$\times$ & E02, E03 & 9,416 \\ \bottomrule
\end{tabular}
\end{table}

\begin{figure*}[t]
    \centering
    \includegraphics[width=\linewidth]{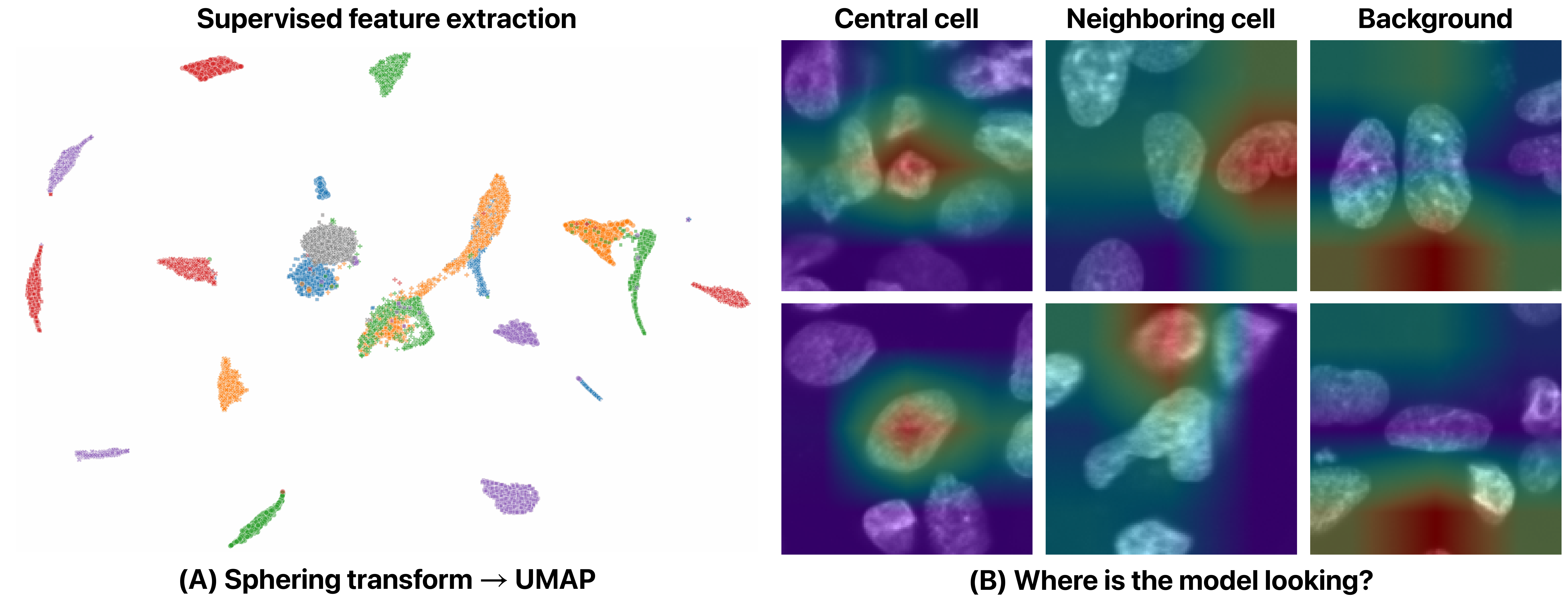}
    \caption{\textbf{(A) UMAP and Grad-CAM.} UMAP embedding of features extracted from the single cells in the test set shows that the different treatment labels are highly separable: \legend. Correction for batch effects is accomplished via a whitening transform. However, cells also cluster by site ($\sbullet[1.5]$, $\mathbf{\times}$, $\ssquare[1.0]$, $\mathbf{+}$), demonstrating the effect of confounding on learned morphological profiles. \textbf{(B) Where is the model looking?} Using Grad-CAM, we identify three patterns in model attention during deep morphological profiling: concentrating on the central cell, a neighboring cell, or the background. 
    In these localization maps, red denotes higher attention. 
    For visualization purposes, only render the central slice of the Z-stack and overlay associated slice in the Grad-CAM localization map.}
    \label{fig:umapcam}
\end{figure*}

Segmentation with Cellpose enumerated 52,311 cells across all biological and technical replicates~(\Cref{tab:dataset}), which were subsequently processed into a dataset of single-cell crops. Partial cells on the boundary of a Z-stack were removed from the dataset and are not included in the total cell counts. To mitigate the potential for data leakage between the training and testing datasets, we used a leave-wells-out validation strategy, training all models on cells from wells in the second column of the plate and evaluated on cells in the third column.
A subset of cells within the training set were used as a held-out validation set for an early stopping criterion.

\begin{figure*}[t]
    \centering
    \includegraphics[width=\linewidth]{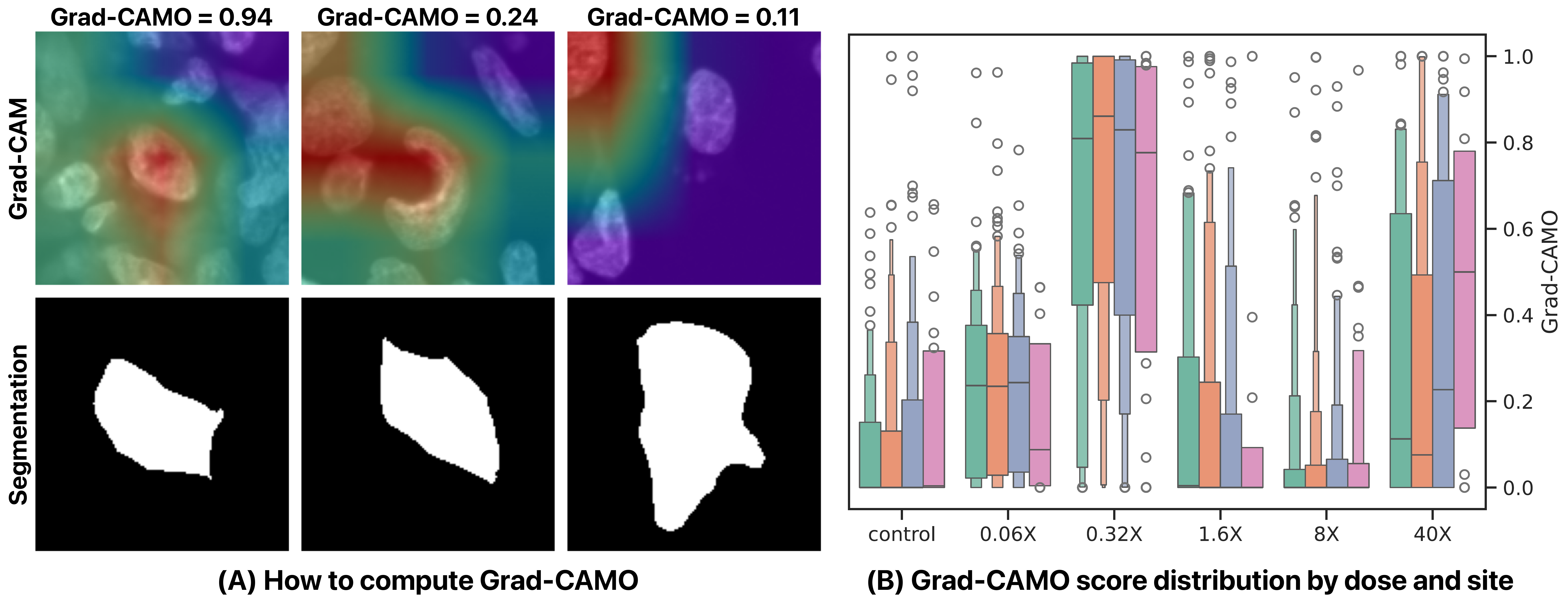}
    \caption{\textbf{(A) Example Grad-CAMO scores.} Grad-CAMO is calculated as the proportion of the model's Grad-CAM localization map that lies within the segmentation mask of the central cell. \textbf{(B) Grad-CAMO scores computed over the entire testing set.} Distributions are grouped by treatment dose and which site in the well was imaged (\ie technical replicate): \legendary.}
    \label{fig:gradcamo}
\end{figure*}

\subsection{Visualizing Learned Morphological Profiles}

A 3D EfficientNet-B0 was trained to predict the treatment label of single-cell crops for ten epochs (or roughly two hours). This short training time was sufficient for both training and validation accuracy to plateau at 92\% and 89\%, respectively, demonstrating the simplicity of this classification task for convolutional models. At the end of each epoch, morphological profiles were extracted from single cells in the test set and visualized using UMAP (\Cref{fig:umapcam}A). Prior to dimensionality reduction, a whitening transform centered on cells in the control group was performed on extracted features in an attempt to eliminate batch effects. Visualization of the learned morphological profiles following batch correction demonstrates that cells in the test set are also highly separable. That is, deep learning strategies extrapolated from 2D use-cases are also able to learn representations of 3D Cell Painting images that are discriminatory based on a categorical variable (here, the model achieves 85\% accuracy on the test set). However, high classification accuracy belies a strong source of confounding in the extracted features: cells in the UMAP cluster not only by the treatment dose, but also by which of the four sites the cell was imaged from (\ie cells also cluster by technical replicate). While the whitening transform removes this source of variation for control cells by mapping their morphological profiles to a standard Gaussian distribution, this batch correction method is insufficient to recover faithful morphological representations for the other treatment doses.

Visualizing Grad-CAM localization maps reveals three attention patterns when the model is predicting the treatment of a previously unseen cell: (1) the model concentrates on the central cell of interest, as intended; (2) the model concentrates on the morphology of a neighboring cell; and (3) the model concentrates on some biologically irrelevant noise in the background of the image. 
These three modes are visualized in \Cref{fig:umapcam}B. 

This experiment illustrates two mechanisms by which the morphological profiles estimated by supervised machine learning models are misaligned with the expectations of human practitioners.
First, we see that models achieve high accuracy by sometimes classifying neighboring cells, perhaps because the neighbors in these cases closely resemble a cell in the training set. However, this is incongruous with our assumption that single-cell feature vectors are describing the central cell in a 3D crop.
Second, this visualization also demonstrates that supervised feature extractors can simply cheat by exploiting non-biological information in microscopy data. In the examples in \Cref{fig:umapcam}B, there are no human-discernible features in the background of the image that would enable the practitioner to determine the treatment received by these cells. While the UMAP embedding demonstrates highly distinct clusters, visual interpretation of these morphological profiles show that deep learning models do not always capture the biological information we expect them to.

\subsection{Computing Grad-CAMO Scores}
Visualizing Grad-CAM localization maps helps diagnose failure modes for supervised feature extractors. However, manually inspecting every morphological profile in this manner would be too labor-intensive. To efficiently evaluate our trained model, we instead compute the Grad-CAMO score for every cell in the test set. 

An illustration of how Grad-CAMO is calculated is shown in \Cref{fig:gradcamo}A.
In the first example, the Grad-CAM localization map is centered on the cell of interest as desired, resulting in a Grad-CAMO score of 0.94. In the second example, the model concentrates on the unusual cellular morphology, noting the absence of a round nucleus. However, the model also concentrates on neighboring cells, resulting in a Grad-CAMO score of 0.24. In the final example, the model only concentrates on a neighboring cell and receives a Grad-CAMO score of 0.11.
Cells in the test set had an average Grad-CAMO score of $0.26 \pm 0.34$. 
Plots of the distribution of Grad-CAMO scores, grouped by treatment dose and site, show that batch effects present in certain treatment dosages confound the morphological profiles extracted by this model (\Cref{fig:gradcamo}B). 

While the model achieves high classification accuracy on the test set and feature vectors visualized via UMAP cluster neatly, analysis with Grad-CAMO demonstrates that these single-cell morphological profiles have low interpretability and poorly represent the central cell in each crop. In fact, we find that only 30\% of learned morphological profiles have Grad-CAM localization maps that significantly overlap with the cell's segmentation map (using a Grad-CAMO score of $0.25$ as a cutoff).

Filtering morphological profiles with low Grad-CAMO scores is a potential strategy for removing additional batch effects. For example, see the UMAP in \Cref{fig:abstract}, which only plots morphological profiles with a Grad-CAMO score greater than $0.25$. However, this removes roughly 13,000 cells from the dataset, reducing the power of the study. Alternatively, Grad-CAMO could be used to evaluate different design choices for a feature extraction pipeline. After determination of an optimal model architecture, low Grad-CAMO can be used to filter outlier cells with unrepresentative morphological profiles.
From a performance point of view, Grad-CAMO is a fast and memory-efficient operation that can be calculated in parallel with the feature extraction step, increasing computational overhead by from \SI{22}{\minute} to \SI{34}{\minute} for 23,000 cells on a single GPU. Computing Grad-CAMO on an entire dataset can also trivially be parallelized across multiple GPUs.

\section{Discussion}

Given the critical importance of single-cell feature extraction for numerous downstream biological tasks (\eg measuring dose response, elucidating mechanisms of action, or predicting clinical outcomes, \etc), ensuring the fidelity and interpretability of morphological profiles is paramount. Deep learning-based feature extraction has many advantages over classical image processing pipelines using manually engineered features.
For example, deep models are more computationally efficient, as they execute on the GPU, and are therefore better suited for high-throughput imaging data. Most importantly, they can learn rich feature representations that describe phenotypes not captured by hand-crafted metrics.
However, the increased expressiveness of black-box models comes at a cost: our experiments identify multiple mechanisms by which the features extracted by neural networks are misaligned with the expectations of human analysts, including describing the morphologies of neighboring cells or simply cheating by exploiting imperceptible noise patterns in the image background. 

Grad-CAMO, our proposed explainability metric, can distill the degree of biological relevance of a morphological profile using a simple and interpretable score.
This score can be evaluated per-sample, to filter unrepresentative morphological profiles, or across an entire dataset, to measure the impact of various architectural design choices on feature extraction pipelines. Maximizing Grad-CAMO serves to optimize the fidelity of morphological profiles to cells of interest while maintaining the superior expressiveness of deep learning models. To this end, Grad-CAMO can be used to align the capabilities of powerful supervised feature extractors with our expectations of how they perform representation learning.

By improving the explainability and interpretability of supervised deep learning models, Grad-CAMO has the potential to improve the utility of single-cell morphological profiles. To conclude, we outline future directions for Grad-CAMO that we are curious to explore.

\paragraph{Grad-CAMO as a heuristic for hyperparameter tuning.}
The design space of feature extraction pipelines is incredibly large, comprising multiple data preprocessing strategies, augmentation options, convolutional backbones, layers for feature extraction, choices of pooling layers, \etc. Oftentimes, design choices are made arbitrarily, and in the absence of a specified downstream task for which there are ground truth labels, evaluation of these choices is difficult. To overcome this challenge, Grad-CAMO can serve as a heuristic to evaluate the impact of different architectural choices.

\paragraph{Grad-CAMO as a regularizer during training.}
It is desirable that the morphological profile extracted by a deep learning model represent the central cell in a 3D crop, instead of a neighboring cell or the background. 
This can be enforced during training by using Grad-CAMO as a regularizer. 
That is, rather than throwing away single-cell segmentation masks after data preprocessing, these labels can be used to compute Grad-CAMO during training. In addition to a classification loss, we can also minimize negative Grad-CAMO computed across all cells. This form of regularization will condition intermediate activations of neural networks to focus on the central cell.

\paragraph{Cell Painting inter-channel correlation.}
A particularly interesting of Grad-CAMO left unexplored in this work is its extension to vision transformers-based models. Unlike convolutional architectures, vision transformers have a built-in mechanism that enables visual explainability: the self-attention map. The self-attention map can be used in place of the Grad-CAM localization map to calculate Grad-CAMO for vision transformers. The current lack of open-source datasets of 3D Cell Painting images hinders the development of fully self-supervised 3D vision transformers for morphological profiling. However, supervised transformer-based models can also exploit the insights gained by through Grad-CAMO to quantitative measure the quality of morphological profiles obtained by this increasingly important class models.




{\small \bibliographystyle{ieeenat_fullname} \bibliography{main}}
\end{document}